\documentclass[twocolumn,showpacs,preprintnumbers,amsmath,amssymb]{revtex4}
\usepackage{graphicx}% Include figure files
\usepackage{dcolumn}% Align table columns on decimal point
\usepackage{bm}% bold math

\DeclareGraphicsExtensions{.eps}

\begin{document}
\preprint{Cerchez et al.}

\title{Absorption of ultrashort laser pulses in strongly overdense targets}
% Force line breaks with \\

\author{M. Cerchez}\email{mirela.cerchez@uni-duesseldorf.de}
\author{R. Jung}
\author{J. Osterholz}
\author{T. Toncian}
\author{O. Willi}

\affiliation{Heinrich-Heine Universit\"at D\"usseldorf, Universit\"atsstr.1,
40225 D\"usseldorf, Germany}
\author{P. Mulser}
\affiliation{Theoretical Quantum Electronics (TQE), Technische Universit\"at Darmstadt, 64289 Darmstadt, Germany}
\author{H. Ruhl}
\affiliation{Institute for Theoretical Physics I, Ruhr-Universit\"at Bochum, 44797 Bochum, Germany}
\date{\today}

\begin{abstract}
Absorption measurements on solid conducting targets have been
performed in s and p polarization with ultrashort, high-contrast
Ti:Sa laser pulses at intensities up to
$5\times10^{16}\,\mathrm{W/cm^2}$ and pulse duration of 8 fs. The
particular relevance of the reported absorption measurements lies in
the fact that the extremely short laser pulse interacts with matter
close to solid density during the entire pulse duration. A
pronounced increase of absorption for p polarization at increasing
angles is observed reaching 77\% for an incidence angle of
80$^\circ$. Simulations performed using 2-D Particle-In-Cell code
show a very good agreement with the experimental data for a plasma
profile of $L/\lambda\approx 0.01$.
\end{abstract}

\maketitle
\indent In the recent years, the availability of intense, ultrashort laser pulses has made the investigation of rapidly heated matter
under extreme conditions possible \cite{Gibbon_Forster, Linde, Saemann, Price}. These interaction processes are characterized by
unique proprieties as the laser energy is absorbed on a short time scale, before significant hydrodynamic motion of the plasma occurs
and the laser energy is transferred to high density matter. The experimental and theoretical studies of the interaction of
femtoseconds laser pulses with solid targets have been motivated by many research fields and applications including ignition methods
for ICF \cite{Tabak}, generation of ultrafast x-rays \cite{Kmetec} and high harmonics \cite{Linde}, highly energetic particles
production \cite{Teubner}, or isochoric heating \cite{Saemann}. Dense plasmas ($n_e>10^{22}\,\mathrm{cm^{-3}}$) generated by
ultrashort laser pulses represent an important interest for astrophysics (e.g. the study of the x-ray opacity of the matter in
similar states as one finds in stars \cite{Rose}), investigations in high-quality laser material processing \cite{Amoruso},
transport properties \cite{Fisher} or dense material equations of state.\\
\indent A central issue of these applications is the question how
and with what efficiency the laser energy is transferred to solid
matter. Absorption of intense laser pulses in the ps \cite{Kieffer}
and sub-ps regimes \cite{Sauerbrey, Feurer, Chen, Price} has been
measured in several experiments with solid targets in the past, in
an intensity range which is of interest to the present letter. In
all measurements, except \cite{Chen} (no s polarization
investigated), absorption has prevailed considerably under p
polarization relative to s polarization. When the laser pulse is
typically longer than 100 fs, and/or a prepulse is present, a
pre-plasma is formed in front of the target and undergoes
hydrodynamic expansion. In these situations, the absorption process
has a characteristic dependence on the polarization and, for p
polarization, is commonly attributed to the mechanism of linear
resonance absorption. Only the authors of \cite{Chen} make an
exception by claiming that vacuum heating \cite{Gibbon} would
dominate on resonant coupling. Under striking incidence and
non-relativistic laser intensities for p polarization absorption
levels as high as $80\%$ have been measured \cite{Sauerbrey}.
Particle-In-Cell (PIC) \cite{Chen, Gibbon, Wilks} and Vlasov
simulations \cite{Ruhl} are in qualitative agreement with
experiments and the characteristic angular distribution of linear
resonance absorption with the increasing scale
length is well reproduced.\\
\indent In this Letter, we report on experimental investigations of laser absorption of high-contrast, sub-10 fs laser pulses by
a conducting target over a large range of angular incidence and laser intensity
$(5\times10^{12}\,\mathrm{W/cm^2}-5\times10^{16}\,\mathrm{W/cm^2)}$. Our laser pulse parameters (duration and high-contrast) allowed,
for the first time, to study the absorption under novel conditions where the pulse energy is basically directly transferred to the
 solid matter. The energy of these extremely short laser pulses can be efficiently absorbed up to $\approx77\%$ by a plasma at
 density close to solid state, characterized by a very steep profile. The absorption of the p polarized laser pulses significantly
 exceeds the s polarization absorption. Computer simulations are consistent with the experimental results for a plasma profile of
 $L/\lambda\approx 0.01$.\\
\indent The experiments have been carried out employing a Ti:Sa
laser system described in \cite{Hentschel} operating in CPA mode.
Under experimental conditions, the laser system delivers linearly
polarized pulses of 100-120 $\mu$J at 790 nm (central wavelength)
and 8 fs duration on target. The pulse contrast was experimentally
determined using a high dynamic range third-order auto-correlator
(Sequoia). The diagnosis reveals a contrast ratio of $10^5$ for
times larger than 1 ps before the main pulse and better than $10^8$
for the Amplified Spontaneous Emission  prepulse. The laser pulse
was focused in vacuum onto target by an f/2.8 off-axis parabola of
108 mm effective focal length to a spot diameter of $\approx$ 3.2 $\mu$m (FWHM), having at normal incidence, an average intensity
of $(4 - 5)\times10^{16}\,\mathrm{W/cm^2}$. The absorbed energy fraction was experimentally determined as a function of the incidence
angle $\theta$, laser pulse polarization and intensity. The targets consisted of mirror-flat aluminium layers with a thickness of
$\approx$ 300 nm and a roughness less than 5 nm, deposited on planar silica substrates. The target was placed at the center of an
integrating sphere of 10 cm in diameter. The amount of laser light collected by the sphere was measured with a high-speed photodiode
coupled to the sphere via an optical fiber bundle. For our energy range, we carefully checked the linearity of the photodetector and
the optical bundle prior to the measurements.\\
\begin{figure}
\includegraphics{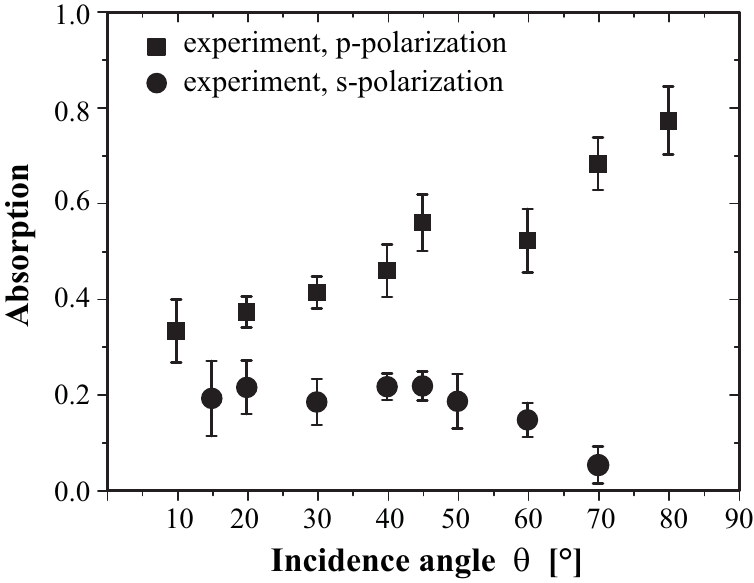}
\caption{Experimental angular dependence of the absorption of 8 fs,
790 nm laser pulses by an aluminium target, s (circle filled
 symbols) and p polarized (squared filled symbols) at an average intensity of $5\cdot10^{16}\,\mathrm{W/cm^2}$.}
\label{Fig.1}
\end{figure}
\indent The laser was operated in single shot mode and focused onto
fresh target surface. The pulse energy fluctuated by less than 5\%
with respect to the average value over tens of shots. The intensity
of the laser beam was varied by moving the target out of focus along
the laser propagation direction. The signal at the photodetector was
proportional to the fraction \textit{R} of the laser energy
reflected (specular and scattered) from the target. Previous
experimental works (e.g.\cite{Borghesi}) proved that the
contribution of the backscattered laser energy represents less than
4\% for incidence angles larger than 10$^\circ$ and thus, is
negligible. The experimental investigations addressed here do not
include the measurement of the backscattered light. The absorbed
fraction \textit{A} is given by $\textit{A}=1-\textit{R}$.\\
\begin{figure}
\includegraphics{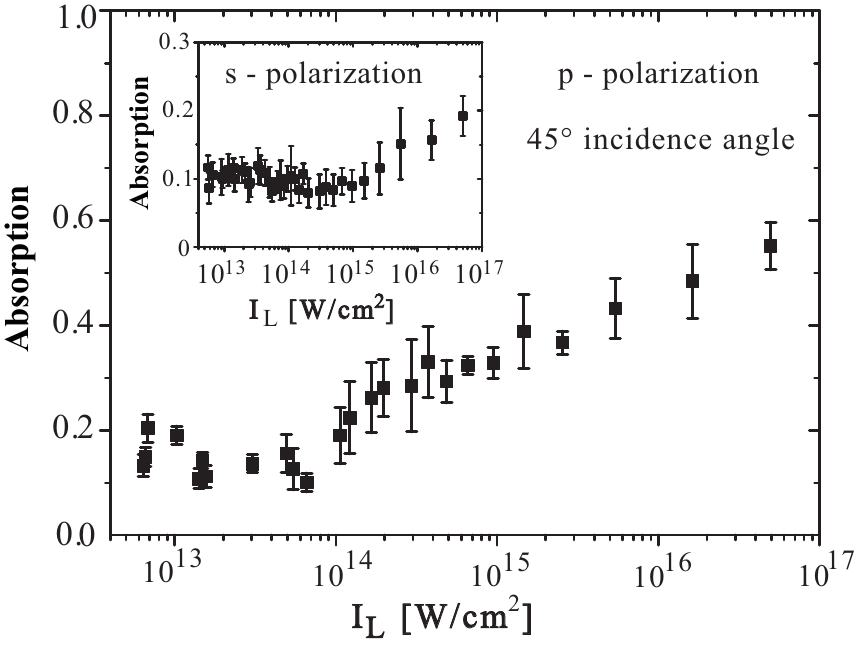}
\caption{Absorption of sub-10 fs, 790 nm laser pulses dependence on
the laser intensity at the incidence angle $\theta = 45^\circ$, for
an aluminium target. In the main frame, the absorbed fraction of the
p polarized beam is shown and in the inset the dependence also on
the laser intensity for a s polarized laser pulse.} \label{Fig.2}
\end{figure}
\indent In Fig. \ref{Fig.1}, the experimental results of the angular
dependence of the absorbed fraction for both s and p polarized laser
pulses incident on aluminium targets are presented. For s
polarization, while the angle of incidence is increasing, the
absorption drops from 19\% at $\theta = 15^\circ$ to $6\%$ at
$\theta = 70^\circ$. Absorption of the p polarized laser light
increases for larger angles and reaches its maximum value of $77\%$
at $80^\circ$. Each data point represents an average value over 10 -
20 shots. The error bars shown on the graphs indicate the standard
deviations and are in the range of $10-40\%$. These experimental
data points have been recorded in the best focal position of the
target. Regarding the dependence of the absorbed fraction versus the
laser intensity ${I_{L}}$, plotted in Fig. \ref{Fig.2} for an
incidence angle of $\theta = 45^\circ$ for both p and s
polarization, the following observations are made: (1) In the case
of s polarization, starting from the low intensity regime of
$\approx5\cdot10^{12}\,\mathrm{W/cm^2}$, the absorption is
approximately constant (10\%) over 2 orders of magnitude of the
laser intensity. It starts to increase up to about 20\% at the
average intensity of $5\cdot10^{16}\,\mathrm{W/cm^2}$. (2)
P-absorption starts to increase significantly at
$10^{14}\,\mathrm{W/cm^2}$ and is 5 times as strong when approaching
$5\cdot10^{16}\,\mathrm{W/cm^2}$. (3) In the intensity range between
$5\cdot10^{14}\,\mathrm{W/cm^2}$ and
$5\cdot10^{16}\,\mathrm{W/cm^2}$ the absorbed fraction of the p
polarized beam $A_{p}$ scales with intensity ${I_{L}}$ and
wavelength as $A_p\propto({I_{L}}\cdot\lambda^2)^{0.12\pm0.02}$.\\
\begin{figure}
\includegraphics{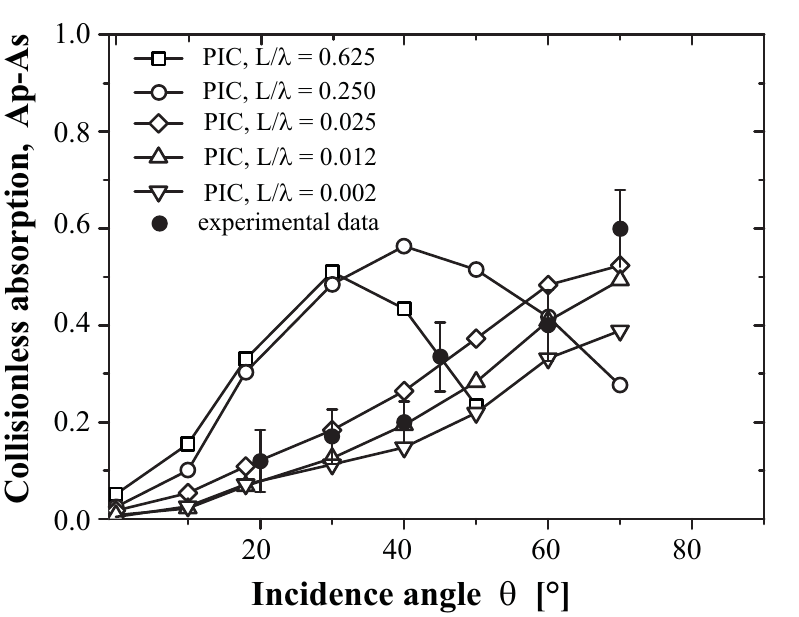}
\caption{Contribution of the collisionless absorption on p
polarization expressed as difference between absorbed fraction on p
and s polarization at an average laser intensity
$I=2\cdot10^{16}\,\mathrm{W/cm^2}$. The filled circle symbols
represents the experimental data. PIC simulations for the same laser
condition and different pre-plasma scale lengths $L/\lambda$ are
shown by open symbols.} \label{Fig.3}
\end{figure}
\indent For investigating these experimental observations, we performed simulations using the 2-dimensional Particle-In-Cell
(PIC) Plasma Simulation Code (PSC) \cite{Ruhl2}. A temporally Gaussian shaped laser pulse with a duration of 10 fs (FWHM) was
focused inside a simulation box of 20 $\mu$m$ \times $20 $\mu$m. In the focus, the laser pulse achieved an intensity of
$2\times10^{16}\,\mathrm{W/cm^2}$ (cycle averaged). A rectangular aluminium target of 15 $\mu$m$\times$1 $\mu$m was placed at
the best focal position in the center of the simulation box. The electron density profile at the target boundary was given by
$n_{e}\left(x\right)/n_{e0}=(1+\exp(-2x/L))^{-1}$ with respect to the target normal, where $L=\left|\nabla n/n\right|^{-1}$ is
the density scale length. For the simulations reported here a grid resolution of 40 cells per $\mu$m was chosen with 4 particles
per cell and species. The target was rotated in steps of $10^\circ$ around the position of the maximum density slope located at
the best focal position (i.e. corresponding to $n_{e}/n_{e0}=1/2)$. Various density scale lengths \textit{L} of 2 nm, 10 nm, 20 nm,
200 nm and 500 nm were used. Additional simulations were performed for higher resolutions up to 120 cells per $\mu$m, but the results
 were not sensitive to the increased resolution. To account for the high contrast-ratio of the laser pulse, it was assumed that the
  target is primarily ionized by optical field ionization at the moment the 10-fs pulse interacts with the aluminium slab. Therefore,
  in the simulation, the aluminium was ionized from the ground state according to the ADK model \cite{Ammosov} included in the code.
  The maximum field-induced ionization state observed in this scenario was about 3.5. The simulations were performed for
  p and s polarization of the incident laser pulse including the binary collisions in the code. Assuming collisional absorption
  for s polarization of the laser, we calculated the contribution of the
collisionless absorption by subtracting the absorbed
  fraction $A_s$ from the corresponding fraction $A_p$ of the p polarized beam. These results have been compared with simulations
  in which the collisional module was switched off and similar results were obtained. This indicates that the fraction $A_p-A_s$
  represents indeed a measure for the contribution of a collisionless absorption process. The computational and experimental results
  at an average focal intensity of $2\cdot10^{16}\,\mathrm{W/cm^2}$ are shown in Fig. \ref{Fig.3}. The experimental data are well
  reproduced for profiles in the range of 10-20 nm. For longer profiles, \textit{L}=200 nm and \textit{L}=500 nm, the well known linear
  resonance absorption behavior is reproduced with an optimum absorption angle at intermediate values.\\
\indent The simulations confirm the absorption of p polarization up
to $80^\circ$ and indicate that the interaction of the ultrashort
laser pulses with the target takes place close to the solid density.
There are more experimental indications, in addition to the above
mentioned good contrast ratio that the laser is incident at steep
plasma density: $\left(i\right)$ in the structure of X-ray spectra
as the higher order transitions are missing \cite{Osterholz} and
$\left(ii\right)$ absence of a preplasma in the observations
of the ionization front propagation in gaseous targets \cite{Jung}.\\
\indent Previous experimental and theoretical works emphasize an
increase and shift of maximum absorption towards larger angles in
linear \cite{Eidmann, Kruer, Landen, Fedosejevs} and nonlinear
\cite{Ruhl} resonance absorption as \textit{L} decreases. We shall
emphasize that the classical model of linear resonance absorption
\cite{Kruer} is not valid for these steep profiles. In such plasmas,
the plasma frequency $\omega_{p}$ is everywhere much higher than the
laser frequency $\omega$. In \cite{Landen, Fedosejevs} it was shown
that in very steep plasma profiles, within a suitable range of
parameters, the resonance absorption approaches the purely
collisional model described by Fresnel equations. As mentioned
above, the PIC simulations indicate that the preponderance of the p
polarization absorption on s polarization found in our experiment
is of collisionless nature. Moreover, the above mentioned threshold behavior is very unlike to be produced in collisional processes.\\
\indent In principle, other collisionless or collective absorption models exist showing a polarization dependence, like sheath layer
inverse bremsstrahlung \cite{Catto}, anomalous skin layer absorption \cite{Yang}, vacuum heating ($VH$) \cite{Chen, Gibbon},
excitation of surface plasmons \cite{Macchi}, and the Brunel effect \cite{Brunel}. Mechanisms such as described in references
\cite{Chen, Gibbon, Catto, Yang} lead to the absorption of laser energy even in the absence of a low density plasma shelf and,
in the non- or weakly relativistic regime, the contributions of these processes do not exceed $5-10\%$ \cite{Bauer}. For example,
assuming a plasma profile of 10 nm then the electron quiver amplitude, $x_{osc}=eE/m\omega_L^2$, exceeds the plasma profile and
consequently, we analyzed in particular the vacuum heating model. In \cite{Gibbon} a detailed analysis of the $VH$ mechanism and
its dependence on the plasma profile, \textit{L} was performed via PIC simulations. For a laser intensity of
$1\times10^{16}\,\mathrm{W/cm^2}$ and a plasma profile of $L/\lambda=0.01$, the contribution of the $VH$ process to the laser
absorption was found to be very small, with a maximum of $\approx10\%$ at $45^\circ$ incidence angle. This value is significantly
 smaller than the absorption measured in our experiment. Moreover, $VH$ mechanism presents a couple of characteristics resulting
 in specific scaling laws which are not identified in our experiment. For example, in the context of the $VH$ model,
 the collisionless absorption fraction scales with the laser irradiance as $A_{VH}\propto({I_{L}}\cdot\lambda^2)^{0.5}$ while
 from our experimental results $(A_p-A_s)\propto({I_{L}}\cdot\lambda^2)^{0.10\pm0.05}$. This mismatch between the specific $VH$
 scaling laws and our experimental results, leads us to the conclusion that vacuum heating is not the dominant mechanism in the
 present experiment. Its presumable small contribution predicted by PIC simulations in \cite{Gibbon} was not possible to be
 distinguished from our experimental results. Presently, new ideas \cite {Mulser} are under discussions which aim to understand
 the physical mechanisms responsible for the high collisionless absorption experimentally observed in overdense plasma regime at
 large incidence angle. These new models will be discussed in a separate publication.\\
\indent In conclusion, we report on the first absorption experiments
of sub-10 fs high-contrast Ti:Sa laser pulses incident on solid
targets. The very good contrast of the laser pulse assures the
formation of a very small pre-plasma and the pulse interacts with
the matter close to solid density. Experimental results indicate
that p polarized laser pulses are absorbed up to $77\%$ at
$80^\circ$ incidence angle. The simulation results of PSC code
clearly confirm the observations and show
that the collisionless absorption works efficiently in steep density profiles with a scale length of $L/\lambda\approx 0.01$.\\
\indent We would like to thank the laser staff at HHU D\"usseldorf
for the assistance during the experiments. This work has been performed within the SFB/Transregio TR 18 and GRK 1203 programs.\\

\end{document}